\documentclass[%
 reprint,% linenumbers,
superscriptaddress,
 amsmath,amssymb,
 aps, %physrev,
pra]{revtex4-2}

\usepackage{silence}
\ErrorsOff*

\usepackage{graphicx}% Include figure files
\usepackage{dcolumn}% Align table columns on decimal point
\usepackage{bm}% bold math
\usepackage{multirow}%
\usepackage{amsmath,amssymb}%
\usepackage{amsthm}%
\usepackage{xcolor}%
\usepackage{textcomp}%
\usepackage{booktabs}%
\usepackage{algorithm}%
\usepackage{algorithmicx}%
\usepackage{algpseudocode}%
\usepackage{listings}%
\usepackage{braket}
\usepackage{xfrac}
\usepackage{subcaption}
\usepackage{lmodern}
\usepackage{anyfontsize}
\usepackage[colorlinks=true, allcolors=blue]{hyperref}

\newcommand{\labelphantom}[1]{%
  \parbox{0pt}{\phantomsubcaption\label{#1}}%
}
\begin{document}

\title{\textbf{Parametric tuning of quantum phase transitions in ultracold reactions} 
}% 

\author{Vijay Ganesh Sadhasivam}\email{vijay13101996@gmail.com}
\affiliation{Theoretical Division, Los Alamos National Laboratory, Los Alamos, 87545, New Mexico, USA}
\affiliation{Yusuf Hamied Department of Chemistry, University of Cambridge, Lensfield road, Cambridge, CB2 1EW, United Kingdom.}

\author{Fumika Suzuki}%\email{fsuzuki@lanl.gov}
\affiliation{Theoretical Division, Los Alamos National Laboratory, Los Alamos, 87545, New Mexico, USA}
\affiliation{Center for Nonlinear Studies, Los Alamos National Laboratory, Los Alamos, 87545, New Mexico, USA.}

\author{Bin Yan}%\email{byan@lanl.gov}
\affiliation{Theoretical Division, Los Alamos National Laboratory, Los Alamos, 87545, New Mexico, USA.}

\author{Nikolai~A. Sinitsyn}\email{nsinitsyn@lanl.gov}
\affiliation{Theoretical Division, Los Alamos National Laboratory, Los Alamos, 87545, New Mexico, USA.}

\date{\today}% It is always \today, today,
             %  but any date may be explicitly specified

\begin{abstract}
Advances in atomic physics have led to the possibility of a coherent transformation between ultracold atoms and molecules including between completely bosonic condensates. Such transformations are enabled by the magneto-association of atoms at a Feshbach resonance which results in a passage through a quantum critical point. In this study, we show that the presence of generic interaction between the constituent atoms and molecules can fundamentally alter the nature of the critical point, change the yield of the reaction and the order of the consequent phase transition. We find that the correlations introduced by this interaction induce nontrivial many-body physics such as coherent oscillations between atoms and molecules, and a selective formation of squeezed molecular quantum states and quantum cat states. We provide analytical and numerical descriptions of these  effects, along with scaling laws for the reaction yield in non-adiabatic regimes.

\end{abstract}

\maketitle
\section{Introduction}
Ultracold atoms hold a prominent place in quantum engineering \cite{bohn2017cold}, information processing \cite{baranov2012condensed}, and further understanding of fundamental quantum phenomena such as superconductivity, magnetism and mechanisms of elementary chemical reactions. One of the landmark experiments in the field is the creation of Bose-Einstein condensates (BECs) using dilute gases of alkali metal atoms \cite{anderson1995observation,DavisKetterle1995, Bradley1995} in the 1990s. A series of developments have led to the creation of a degenerate Fermi gas \cite{de2019degenerate}. More recently, it became possible to observe a {\it quantum coherent chemical reaction} for a macroscopic number of bosonic atoms forming a degenerate Bose gas of molecules \cite{zhang2021transition,zhang2023many}.

The association of bosonic/fermionic atoms into molecules is achieved by tuning a time-dependent magnetic field across the {\it Feshbach resonance}, which drives an ultracold atomic system through a quantum phase transition \cite{romans2004quantum,radzihovsky2004superfluid}. This `reaction' is quantum coherent and reversible at macroscopic scale. 
The \textit{time-dependent Tavis-Cummings (TC) model} \cite{yurovsky2002quantum} is a minimal model that captures many non-trivial many-body effects that are encountered during the stimulated reaction. Its Hamiltonian is 
\begin{align}\label{H_TC}
\hat{H}_{TC}(t) =-\beta t \hat{\psi}^{\dagger} \hat{\psi} + \frac{g}{\sqrt{N}}(\hat{\psi}^{\dagger} \hat{S}^{-}+\hat{\psi} \hat{S}^+), \quad N\equiv 2S.
\end{align}

Here, $S$ ($\gg1$) is the quantum mechanical {\it pseudo-spin}, and $\beta$ is the {\it sweep rate} of the transition through the Feshbach resonance. The sweep rate is experimentally controllable in a broad range -- from almost instantaneous to quasi-adiabatic.  The bosonic   operator, $\hat{\psi}^{\dagger}$, creates a molecule. The terms with the pseudo-spin raising and lowering operators,  $\hat{S}^+$ and $\hat{S}^-$, correspond, respectively, to the dissociation of a molecule and association of two \textit{fermionic} atoms to form a bosonic molecule  with a characteristic coupling constant $g$. The driven TC model can also be reformulated in terms of a fully bosonic
reaction between bosonic atomic and molecular condensate (see \cite{altland2009nonadiabaticity,malla2022coherent} and section~\ref{TC-bos} for derivation of this model from true atomic-molecular Hamiltonians). We will focus here on the reaction in which there are no molecules initially. This corresponds in (\ref{H_TC}) to the initial state, as $t\rightarrow -\infty$, without molecules and the spin fully up-polarized along the $z$-axis.

The time-dependent field sweep is needed in practice to make all atoms encounter the resonance. In addition, the quantum adiabatic theorem guarantees that a sufficiently slow sweep converts the initial atomic state into the bosonic ground state, which is the molecular condensate. Hence, potentially, a 100\% efficiency of the reaction is experimentally possible. However, the true adiabatic limit cannot be reached, so it is  important to understand the quasiadiabatic regime with a small but finite $\beta$. This regime is characterized by the number of {\it nonadiabatic excitations}, $n_{ex}$, which is the number of the molecules that would appear in the adiabatic limit but did not form after the transition through the resonance, i.e., as $t\rightarrow +\infty$. 

An unusual theoretical finding about the time-dependent TC model was the discovery of its integrability \cite{sinitsyn2016solvable, sun2016landau}, which confirmed  semiclassical predictions for the power-law scaling of the nonadiabatic excitation density after the quasi-adiabatic sweep through the resonance \cite{altland2009nonadiabaticity}. 

However, recent experiments \cite{zhang2023many} with a bosonic reaction show a behavior that is not known within the slowly driven TC model, such as coherent oscillations between atoms and molecules after crossing the Feshbach resonance. Moreover, even with the slowest sweep rates, a system usually ends up in a prethermalized atomic-molecular mixture state, in disagreement with the nearly perfect reaction efficiency predicted by the TC model \cite{chin2010feshbach}.

One possibility to extend the TC model to account for the richer reaction dynamics is to add a dispersion of the atomic states. This results in a generalized TC model \cite{sun2016landau,malla2022coherent} that, surprisingly, is also solvable and leads to essentially the same predictions for the dependence of the number of nonadiabatic excitations on the sweep rate as the minimal model (\ref{H_TC}). Hence, other interaction types may be responsible for the experimentally observed behavior beyond the standard TC model.

\begin{figure*}
    \centering
     \labelphantom{fig1a}%
    \labelphantom{fig1b}%
    \labelphantom{fig1c}%
    \labelphantom{fig1d}
     \labelphantom{fig1e}%
    \includegraphics[width=\textwidth]{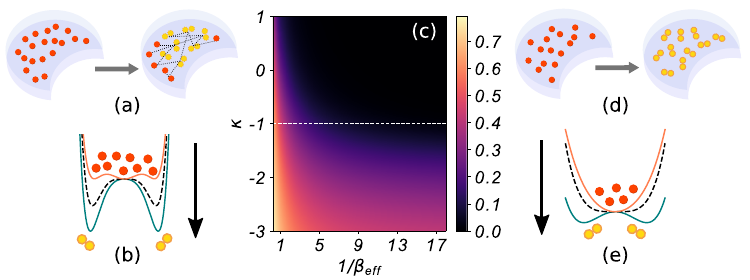}
    \caption{ An ultracold atomic reaction undergoing a (non-adiabatic/quasi-adiabatic) passage through, respectively, (a),(b) the first-order  and (d),(e)  the second-order quantum phase transition. (a) For the first-order transition, both molecules and atoms can be in the local energy minima simultaneously, whereas (d), for the second-order transition, a high efficiency of the chemical reaction can be achieved by  an adiabatic transition between the molecular and atomic energy minima.
    (b) and (e) The mean-field ground energy as a function of the order parameter in the first-order and second-order quantum phase transition, respectively (arrows indicate the direction of time) (c) The numerically obtained phase diagram describing the dependence of the number of nonadiabatic excitations, $n_{\text{ex}}$, as a function of the inverse  sweep rate, $1/\beta_{\rm eff}\equiv g^2/(\beta \ln N)$ ($x$-axis), and the molecular interaction strength, $\kappa=r/g$ in Eqs.~(\ref{H_TC})~and~(\ref{Hint}), ($y$-axis). For the  model (\ref{Htotal}), $\kappa=-1$ separates the regimes with first ($\kappa<-1$) and second ($\kappa>-1$) order phase transitions. This explains the fast increase of the excitation numbers below the $\kappa=-1$ line.}
    \label{fig1}
\end{figure*}
%%%%%%%%%%%%%%%%%%%%%%%%%%%
In this article, we consider a different generalization of the model (\ref{H_TC}), in which we add another interaction term: 
\begin{eqnarray}
\label{Htotal}
&&\hat{H}= \hat{H}_{TC}+\hat{H}_{\rm int},\quad \quad {\rm where}\\
    \label{Hint}
 &&   \hat{H}_{\text{int}} = \frac{r}{N} (\hat{\psi}^{\dagger} \hat{\psi})^2 = \frac{r}{N}\hat{n}^2 ,
\end{eqnarray}
and where $\hat{n}$ is the molecular number operator; $r$ is the interaction strength. This interaction term includes all the `non-resonant' interactions that are possible due to elastic scattering between atoms and molecules and is hence expected to have the same level of significance as the resonance term in \eqref{H_TC}. {Specifically, this interaction term consists of two-body atom-atom, atom-molecule and molecule-molecule scattering terms, all of which can be accounted for using the molecule number operator $\hat{n}$ due to the fact that quantum dynamics conserves the number of particles (explained in detail in sections \ref{sc_sec21} and~\ref{TC-mol})} .  The coupling in (\ref{Hint}) effectively broadens the molecular dispersion energy, which becomes now dependent on the number of molecules that changes during the reaction. Such a coupling must be present in ultracold molecules due to elastic scattering and dipole interactions, which in certain atomic systems can be stronger than the coupling induced by the Feshbach resonance \cite{dipole-1,dipole-2}.

The main reason for the inclusion of the interaction term in \eqref{Hint} is the fact that the TC model (\ref{H_TC}) is a phenomenological model that was introduced as a simplification of a system of bosonic atoms in a confinement with complex pairwise atomic interactions. The full many-body Hamiltonian includes only the interacting atoms in a trap. To obtain the TC model from that Hamiltonian, one should identify a quasi-stable bound state (the molecule), and then keep only the interactions of this bound state with free atomic states (see sections~\ref{TC-bos} and~\ref{TC-mol}). This truncation of
the original Hamiltonian is justified \text{not} by the dominance of the coupling $g$ over all other interactions but rather by the fact that the creation of the molecules from the atoms is a distinct effect, whereas the remaining interactions, the most relevant of which is given by Eq.~(\ref{Hint}), are usually assumed to just renormalize the parameters of quasiparticles in a condensed matter. 

We will show, however, that 
such an omission is not justified when considering a transition through a critical point, particularly when the characteristic energy scales for $r$ and $g$ are comparable. The addition of $\hat{H}_{\rm int}$ leads to substantially richer behavior, especially for quasi-adiabatic transitions. This could be anticipated from the applications of the TC model in optics, where such a nonlinear term describes the optical mode in a `Kerr-like' medium \cite{buvzek1990dynamics, agarwal1989collapse, gora1992nonlinear}.

With this addition, the phase diagram for the reaction efficiency depends on both the sweep rate $\beta $ and the nonlinearity $r$. 
In Fig.~\ref{fig1c} we show the result of our numerical simulations for the number of the nonadiabatic excitations, which is the number of unformed molecules, following the sweep of the chemical potential. We set $r=\kappa g$, where the line $\kappa=-1$ marks a critical nonlinearity; for $\kappa<-1$ the number of the nonadiabatic excitations is much larger than for $\kappa>-1$, especially in the quasi-adiabatic limit (large $1/\beta_{\rm eff}\equiv g^2/(\beta\log_e N)$). 
We will show that this behavior follows from the possibility of either a second- or first-order quantum phase transition during the sweep through the Feshbach resonance. Our model is no longer exactly solvable but in the quasi-adiabatic limit it can be studied analytically in detail. We propose our model as a guide for future experiments that explore the complex quantum dynamics of
interacting molecular condensates. In particular, we will derive the formulas for the efficiency of the reaction between molecular and atomic condensates as a function of the sweep rate $\beta$, which can be also used to estimate the parameters $r$ and $g$ experimentally (see also ~\cite{timmermans1999feshbach,Juha2004collective,liu2008role,naniyil2022observation, duine2004atom, wang2024dynamics}). We also show that our approach can be extended to more complex types of Feshbach resonance, such as the one considered in the experiment described in \cite{zhang2023many}.

\section{Results}

The experimentally relevant regime corresponds to a large number of starting atoms, $N=2S \gg 1$, that can potentially create a large number of molecules during the sweep through the resonance. This forms the basis of our semiclassical approximation, which we develop as follows. First, we rewrite the Schr\"odinger equation for the nonlinear TC model \eqref{Htotal} in the form of the Schr\"odinger equation of a particle in a one-dimensional potential. The large $N$-limit can then be understood by considering the classical equations of motion in the same potential. Next we show that the potential energy changes with time: for the parameter regime $r+g>0$, it starts off as a harmonic oscillator potential as $t\rightarrow -\infty$, which is converted into a double-well potential during the passage through a \textit{second-order} phase transition. The number of the nonadiabatic excitations generated during this passage is obtained from the change of the \textit{adiabatic invariant} upon this transition.

This step reveals the universality of the equations of motion near the critical point for $r+g>0$, so we extend the known exact formula for the TC model at $r=0$ to the case with $r\ne 0$. We then show that for sufficiently strong attractive interactions (i.e. $r+g<0$), there is a qualitative change in the behavior of the nonlinear TC model, due to the emergence of an underlying first-order phase transition. In this case, the potential energy changes from a `triple well' potential before the transition to a double well potential after the transition. We derive scaling relations for the number of nonadiabatic excitations which do not vanish even in the adiabatic limit, which is a characteristic of the first-order transition. We then demonstrate some of the dynamical features accompanying this transition, such as oscillatory dynamics of the excitations and formation of nonclassical states, using numerical simulations.

\subsection{Semiclassical description and phase transitions}\label{sc_sec21}

Note that our model conserves 
\begin{align}\label{cons-1}
    \hat{N} \equiv \hat{\psi}^{\dag}\hat{\psi} + (S + \hat{S}_z),
\end{align}
whose eigenvalue $N=2S$ corresponds to our initial conditions, where the starting state  is the one without molecules. Hence, it is convenient to mark all states by the number of molecules: 
$$
|n\rangle \equiv |n\rangle_m \otimes |S-n\rangle_{S_z},
$$
where $|n\rangle_m$ is the Fock state with $n$ bosonic molecules and $|S-n\rangle_{S}$ is the pseudospin state with z-axis projection $S_z=S-n$. For example, $\hat{\psi} |n\rangle_m = \sqrt{n} |n-1\rangle_m$, $S_z|S-n\rangle_{S} = (S-n)|S-n\rangle_S$, $S^+|S_z\rangle_S =\sqrt{S(S+1)-S_z(S_z+1)}|S_z+1\rangle_S$ and so on. {Note also that due to the conservation law (\ref{cons-1}), the addition of the interatomic interactions of the form $\sim S_z^2$ and atom-molecule (non-resonant) interactions of the form $nS_z$ to the original Hamiltonian would be equivalent to a redefinition of the coupling $r$ in the term (\ref{Hint}) and a time shift (see section~\ref{TC-mol}). We assume that this redefinition is already included in our definition of $r$.}

The initial state as $t\rightarrow -\infty$ corresponds to $|n\rangle =|0\rangle$. The matrix elements $H_{nm}\equiv \langle n| \hat{H}_{TC}+\hat{H}_{\rm int} |m\rangle$ in this basis are given by 
\begin{equation}\label{H_TC_matrix}
\begin{split}
    H_{nm}=& \: (-\beta t n+\frac{r}{N}n^2)\: \delta_{n,m} +  g n \sqrt{\frac{N-n+1}{N}} \:\delta_{n,m-1} \\&+  g(n+1) \sqrt{\frac{N-n}{N}} \:\delta_{n,m+1}
\end{split}
\end{equation}
We will show that the critical point appears at conditions $1\ll n \ll N$, so by disregarding terms of order $(n/N)^3$ and $1/N$, the matrix elements can be written as
\begin{equation}\label{H_TC_matrix-2}
\begin{split}
 H_{nm} =& \: (-\beta t n+\frac{r}{N}n^2)\: \delta_{n,m} +  \\ &g n \left(1-\frac{n}{2N} \right) (\delta_{n,m-1} +  \delta_{n,m+1}) +O(1/n).   
\end{split}
\end{equation}

A formal solution to the time-dependent Schr\"odinger equation for the driven nonlinear TC model \eqref{Htotal} can be written as: 
$|\psi \rangle = \sum_{n=0}^N a_n(t) |n\rangle$. We introduce the {\it amplitude generating function}
\begin{align}
    u(\phi,t ) =  \sum_{n=0}^N a_n(t) e^{in\phi},
    \label{agf-def}
\end{align}
whose name follows from the fact that the amplitudes can be obtained by the inverse Fourier transformation
$$
a_n(t)=\frac{1}{2\pi} \int_0^{2\pi} d\phi e^{-in\phi} u(\phi,t).
$$

For the Hamiltonian (\ref{H_TC_matrix-2}), the time-evolution of the amplitudes, up to the omitted terms $\sim 1/n$, is given by 
\begin{equation}
i\dot{a}_n=(-\beta t n+\frac{r}{N}n^2)a_n +  g n \left(1-\frac{n}{2N} \right)(a_{n-1}+a_n).
\label{amp-ev}
\end{equation}
Note that $n a_ne^{in\phi} = -ia_n\partial e^{in\phi}/\partial \phi$. We then multiply Eq.~(\ref{amp-ev}) by $e^{in\phi}$ and sum all equations for $n=0,1,2,\ldots$. Using the definition in (\ref{agf-def}),
the time-dependent Schr\"odinger equation for the amplitudes is then written in terms of a single equation for $u(\phi,t)$ as 
\begin{equation}
\label{se-u}
i\frac{\partial }{\partial t} u(\phi,t)= \hat{H}\left(-i\frac{\partial}{\partial \phi}, \phi \right) u(\phi,t ),
\end{equation}
where we associate  $\hat{n} \equiv -i\partial/\partial \phi$. In the semiclassical approximation we can then associate $\phi$ with a coordinate that is conjugate to the classical momentum $n$. Then, the classical Hamiltonian that corresponds to the 
Schr\"odinger equation (\ref{se-u}) has the form
\begin{align}\label{H_MF}
    H_{cl}(n,\phi) = -\beta t n +\frac{r}{N} n^2+2gn\left(1 -\frac{n}{2N} \right) \cos(\phi), 
\end{align}
and the classical equations of motion are given by
\begin{equation}
\frac{d \phi}{dt} = \frac{\partial H_{cl}}{\partial n}, \quad \frac{d n}{dt} =-\frac{\partial H_{cl}}{\partial \phi}.
    \label{cl-em1}
\end{equation}

\subsection{Quasi-adiabatic second-order phase transition }
Without the new $r$-dependent nonlinear term, the Hamiltonian (\ref{H_MF}) would coincide with the analogous semiclassical Hamiltonian in \cite{altland2009nonadiabaticity}.
The relation between the classical variables and the number of nonadiabatic excitations is established by noting that as $t\rightarrow \pm \infty$ 
the time-dependent term completely dominates: $H_{cl}\sim -\beta t n$. Following \cite{altland2009nonadiabaticity}, we note that
the equations of motion initially conserve $n$ and the adiabatic invariant is given by the 
initial number of molecules, $n_{-\infty}$:
$$
I_{-} =\frac{1}{2\pi} \int_{0}^{2\pi} n\, d\phi = n_{-\infty}.
$$
If during the evolution the adiabatic invariant acquires a small contribution $\Delta I$, this is interpreted in the semiclassical theory as the density of the produced elementary nonadiabatic excitations, i.e, $n_{ex}=\Delta I/N$. In the strict adiabatic limit, the ground state with no molecules eventually transitions into the new ground state with $N$  molecules. Taking the nonadiabatic excitations into account, the number of molecules created in our process is given by 
$$
n_{+\infty}=N-\Delta I - I_{-}.
$$

We should assume that initially $n_{-\infty}=I_{-\infty}\sim 1$, which is negligible in comparison to the large $N$. 
We assume a nearly adiabatic sweep of the chemical potential \cite{altland2009nonadiabaticity}. The point $\phi^*=\pi$ is a steady point of the classical equations (\ref{cl-em1}), in the vicinity of which the system evolves. Assuming that near this point $n \ll N$, and  retaining the terms of the lowest order we find an effective Hamiltonian that governs the evolution at the early stage of the reaction, that is
\begin{align}
\label{H_MF_IIPT}
     H_{cl}(n,\phi;\gamma) \approx \left( -\beta t - 2g\right) n + \left(\frac{r+g}{N}\right)n^2 + g n\phi^2.
\end{align}

For a quasi-adiabatic evolution, the nonadiabatic excitations are generally suppressed exponentially, as $\Delta I \sim \exp(-a/\beta)$, with some finite positive $a$. Such excitations are not essential and we can safely disregard them. However, according to the Kibble-Zurek phenomenology, the excitations are enhanced near a critical point at which the symmetry of the original ground state breaks down spontaneously. It turns out that the Hamiltonian (\ref{H_MF_IIPT}) contains this event, so it is sufficient to describe the phase transition quantitatively.

\begin{figure}
    \centering 
    \includegraphics[width=\linewidth]{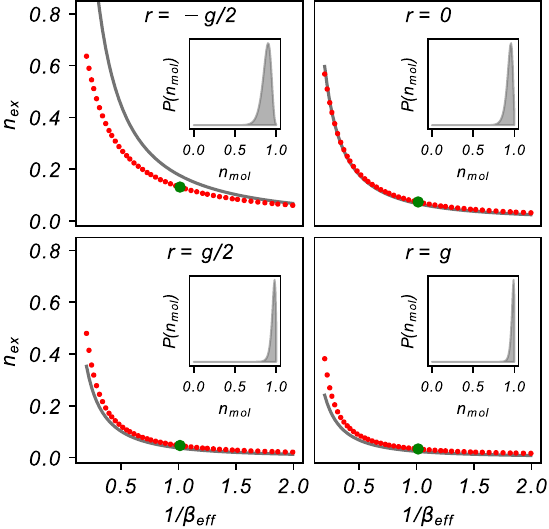}
    \caption{Dependence of the final density of excitations, $n_{\text{ex}}=(N-n_{+\infty})/N$,  on the sweep rates for different values of $\kappa$ in the case of a second-order transition. Insets: Distribution of molecule number in the limit $t\rightarrow\infty$ at the value of $\beta_{\text{eff}}$ marked in green.}
    \label{fig2}
\end{figure}

We  shift the zero of time by setting:
\begin{align}\label{timeshift}
    s = t + 2g /\beta,
\end{align}
and make a canonical transformation in (\ref{H_MF_IIPT}):
\begin{align}\label{can_trans}
n=Q^2, \quad \phi=-P/(2Q),
\end{align}
%which can be inverted as $ Q= \sqrt{n}, \quad P=-2\sqrt{n}\phi$,
leading to an effective classical time-dependent Hamiltonian for a single degree of freedom:
\begin{align}
   \label{eff-H2}
    H_{\rm eff}^{\text{II}}(Q,P,s) := -\beta s \: Q^2 + \left(\frac{r+g}{N}\right)Q^4 +  \left(\frac{g}{2}\right)\frac{P^2}{2},
\end{align}
where now $P$ is treated as the momentum and $Q$ as the coordinate. 

Suppose that $r+g>0$. This corresponds to either repulsive ($r>0$) or weakly attractive ($-g<r<0$) interactions between the formed molecules.
If $s$ were a constant, then the Hamiltonian (\ref{eff-H2}) would describe a nonlinear oscillator with the potential energy
\begin{equation}
    V(Q)=-\beta s \: Q^2 + \left(\frac{r+g}{N}\right)Q^4.
    \label{poten}
\end{equation}
The initial conditions correspond to $Q\sim  1$. In fact, a quantum mechanical treatment of the initial conditions requires the averaging of the behavior over a distribution of small initial values with $0< Q^2\ll N$ \cite{altland2009nonadiabaticity}. However, we will show later that this information is irrelevant within the leading-order semiclassical description. 
The system with the potential (\ref{poten}) experiences the second-order phase transition at $s=0$. Indeed, for $s<0$, the potential $V(Q)$ has a single minimum at $Q=0$ but for $s>0$, there are two local minima at $Q_{min}\sim \pm \sqrt{\beta N s/[2(r+g)]}$. 

In Fig.~\ref{fig1b}, we illustrate that for $s<0$ the system is initially near a single minimum but it follows one of the new minima for $s>0$.
In phase space, this corresponds to crossing a separatrix, in whose vicinity the classical adiabatic invariant is no longer conserved. Thus, so far our approximations were justified because they capture the main source of the nonadiabatic excitations near the phase transition.

The  evolution equations for $Q$ and $P$ with the Hamiltonian \eqref{H_MF_IIPT}, acquire a universal form after rescaling of the variables:
$$
s\rightarrow \lambda s, \quad Q\rightarrow u Q, \quad P \rightarrow v Y,
$$
and
$$
H_{\rm eff} \rightarrow \frac{\lambda}{uv} H_{\rm eff}(\lambda s,uQ,vP),
$$
where $\lambda$, $u$, and $v$ are constants. 
We choose them so that
$H_{\rm eff}$ in terms of the rescaled variables has the form with only numerical coefficients
\begin{equation}
    H=-s \frac{Q^2}{2}+\frac{Q^4}{2}+\frac{P^2}{2}.
    \label{HP2}
\end{equation}
In the new variables, the equations of motion have the canonical form of the Painlev\'e-II equation
\begin{equation}
\frac{d^2Q}{ds^2}=sQ-2Q^3.
\label{P-2}    
\end{equation}

The dynamics does not depend anymore on the relative values of the parameters $r$ and $g$. However, we reiterate that at $r=0$, the model is exactly solvable, so all known facts about the solution of the driven TC model can now be applied to the dynamics according to Eq.~(\ref{P-2}).

The rescaling of variables, however, is not canonical, so it does not conserve the action:
$$
\frac{1}{2\pi}\int_{0}^{2\pi} n \,d\phi=\frac{uv}{2\pi} \oint P\, dQ.
$$
Hence, if $I$ is the adiabatic invariant in the original variables, $n$ and $\phi$, then in the rescaled  $Q$ and $P$ coordinates, the invariant is given by
$ {\cal I}= I/(uv)$. 
For our case, we found
\begin{equation}
   uv = \frac{2N\beta}{g^2 +gr}.
   \label{transition}
\end{equation}

Equation~(\ref{transition}) can be used to establish the relation between the exactly solvable case at $r=0$ and our more general model for $r>-g$.  Namely, for $r=0$, the scaling for the number of the excitations was found to be a power law with a logarithmic prefactor \cite{sun2016landau} (for earlier but only semiclassical derivations of this formula at $r=0$, see also discussion around Eq.~(11) in \cite{Gurarie-PhysRevA} and Eq.~(15) in \cite{Itin}): 
\begin{equation}
\Delta I_{r=0} =\frac{uv}{4\pi} \ln \left(\frac{uv}{4I_{-}\pi}\right)=\frac{N\beta}{2\pi g^2} \ln \left( \frac{N\beta}{2\pi g^2I_{-}}\right),
\label{delI11}
\end{equation}
where either the exact solution or further semiclassical analysis can be used to fix $I_{-}\approx 1$ for our initial conditions in the model (\ref{H_TC}). According to \cite{sun2016landau}, this result is valid only in the quasi-adiabatic case, namely for $1/\ln N \gg \beta/(2\pi g^2) \gg  1/N$.  For example, at $I_{-}=1$, in this range Eq.~(\ref{delI11}) coincides with the exact expression for the position of the maximum of the probability distribution of the excitation number. 

Using (\ref{transition}), and the fact that all $r>-g$ values lead to the universal Eq.~(\ref{P-2}), we can now extend the result in (\ref{delI11}) to the case with arbitrary $r>-g$ by properly rescaling the action variables:
\begin{equation}
\Delta I (r) =\frac{N\beta}{2\pi g(g+r)} \ln \left( \frac{N\beta}{2\pi g(g+r)I_{-}}\right),
\label{delI}
\end{equation}
where $I_{-}$ is the parameter of order $1$ that characterizes the initial conditions.   As this unknown factor appears inside the logarithm in (\ref{delI}), its precise value is not important because its relative contribution to (\ref{delI}) vanishes in the limit $N\rightarrow \infty$ for the quasi-adiabatic evolution with  $1/\ln N \gg \beta/(2\pi (g^2+rg)) \gg  1/N$. For comparison with numerical results, we set $I_{-}=1$. 

Thus, Eq.~(\ref{delI}) extends the expression known in the exactly solvable case, for $r=0$, by merely re-scaling the coupling as 
\begin{align}
    g^2 \rightarrow g^2+gr
    \label{rescale}
\end{align}
for a general value of $r>-g$. In terms of the parameter $\kappa=r/g$, this corresponds to the rescaling $g^2 \rightarrow g^2(1+\kappa)$. 
The smaller sweep rates, $\beta/(g^2+rg) <1/N$, are not captured by the formula (\ref{delI}) because this regime corresponds to the onset of the truly adiabatic evolution, with exponentially suppressed excitations. For experimentally relevant values, $N>10^4$, this truly adiabatic regime cannot be reached, so we leave it without further discussion.

The change of the adiabatic invariant is interpreted in terms of the density of the nonadiabatic excitations: $n_{ex}=(N-\langle n\rangle)/N=\Delta I/N$, which in turn is related to the deviation of the average number of the formed molecules from its maximal value. Our analysis so far 
has been restricted to the quasi-adiabatic regime. 
A comparison with numerical results is shown in Fig.~\ref{fig2}. We indeed find agreement with the theory for the interval $1/\ln(N) \gg (g^2+rg)/\beta \gg 1/N $. However, beyond this interval, the deviations from our formulas are strong. 
Note also that these deviations persist to smaller $\beta$ when $\kappa$  approaches the value $\kappa = -1$, indicating the breakdown of our analysis for $\kappa \le -1$.

In summary, our theory predicts the robustness of the dynamics during a passage through a second-order quantum phase transition for repulsive and moderate attractive interactions ($\kappa>-1$). Its signature in experiments can be the scaling of the density of excitations (unformed molecules) for different sweep rates in the quasi-adiabatic regime: 
$$
n_{ex}\sim \beta \ln \beta, 
$$ which is the same as in the exactly solvable case at $\kappa=0$.

\subsection{First-order phase transition}
The energy~\eqref{H_MF_IIPT} has no finite global minimum for $r<-g$, so we must include the next order term, $\propto n^3$ in the Hamiltonian \eqref{H_TC_matrix}. After the transformations \eqref{timeshift}  and \eqref{can_trans} in the classical limit, we find
\begin{equation}\label{H_MF_IPT}
\begin{split}
H_{\rm eff}^{\text{I}}(Q,P,s) := -\beta s \: Q^2 + \left(\frac{r+g}{N}\right)Q^4 \\ + \left(\frac{g}{4N^2} \right) Q^6 + \frac{g}{2}\frac{P^2}{2}
\end{split}
\end{equation}
Disregarding the higher-order terms in $P$ and $Q$ is justified when the main nonadiabatic effects occur for $1\ll Q^2\ll N$. This requires that 
$$
|\delta r| \ll |g|, \quad {\rm where} \quad \delta r := r+g < 0.
$$
  In addition, we should assume the same initial conditions as in the previous case, $Q(-\infty)=O(1)$.  Consider now only the potential energy
$$
V(Q)= -\beta s \: Q^2 - \left(\frac{|\delta r|}{N}\right)Q^4 + \left(\frac{g}{4N^2} \right) Q^6.
$$
As $s\rightarrow -\infty$, it has a single energy minimum at $Q=0$. With time, two additional energy minima initially appear at higher energy but eventually become the global minima of $V(Q)$. However, the transition into them, for some time, is classically forbidden due to the energy barriers.

By approaching the time moment $s=0$, the minimum at $Q=0$ becomes initially a false vacuum, and for $s>0$ it becomes an unstable local maximum. The steady point at $Q=0$ can then be perturbed by any quantum fluctuation, so at $s=0$ the system has to fall towards one of the remaining minima, which, at this moment are given by
 $$Q_{\pm}=\pm \sqrt{8|\delta r| N/(3g)};$$ with the corresponding energy $V(Q_{\pm})=-64 |\delta r|^3N/(27g^2)$. The system can choose to be in one of the two minima with equal probability. At this point, let us consider that the system escapes towards $Q_+$.

 At $s=0$, the energy of the false vacuum state is $E=0$. Hence, along the escape trajectory  the momentum is given by
 \begin{equation}
    P(Q)=\frac{Q^2}{N}\sqrt{4|\delta r| N/g -   Q^2}. 
     \label{momentum1}
 \end{equation}
The turning points of this trajectory are at $Q_0=0$ and $Q_1=2\sqrt{|\delta r| N/g}$ and its adiabatic invariant is 
\begin{equation}
   I_{\rm fin} = \frac{2}{2\pi}\int_{Q_0}^{Q_1} P(Q) \, dQ = \frac{ N |\delta r|^2}{g^2}.
    \label{I-1st}
\end{equation}
Since the initial value of the adiabatic invariant is $O(1)$, to leading order in $N$, the density of excitations is associated with the invariant (\ref{I-1st}):
\begin{equation}
    n_{ex}^{\text{sat}} = n_{ex}^{\text{sat}} (\delta r\slash g)=\frac{I_{\rm fin}}{N} =\frac{|\delta r|^2}{g^2},
    \label{nex-sat}
\end{equation}
which is \textit{independent} of the sweep rate $\beta$, as long as we consider the quasi-adiabatic dynamics and disregard the exponentially suppressed quantum tunneling events. Thus, the first-order phase transition is associated with a finite density of the excitations even in the quasi-adiabatic regime. We note also that an almost perfect efficiency of ultracold chemical reactions has so far not been achieved experimentally even for the slowest sweeps through the Feshbach resonance. This may be an indication of the presence and importance of molecular interactions of the type that we have considered here.

Equation~\eqref{nex-sat} explains the sharp transition in $n_{\text{ex}}$ observed around $\kappa =-1$ in Fig.~\ref{fig1c},  in the quasi-adiabatic regime.  Qualitatively similar saturation of the number of excitations was found previously in the nonlinear Landau-Zener model \cite{liu2002theory}, in which  this behavior was not related to a spontaneous symmetry breaking. The dependence of $n_{\text{ex}}$ on $\beta$ for different values of $r$ with $r<-g$ (or $\kappa<-1$) are plotted in Fig.~\ref{fig3a}, confirming the saturation of $n_{ex}$ at a finite value in the adiabatic limit.

The $\beta$-dependent correction to Eq.~(\ref{nex-sat}) deserves  a discussion.  For small $\delta r$, this correction can be considerable because the saturation value of $n_{ex}$ in (\ref{nex-sat}) scales on $\delta r$ as  $n_{ex} \propto (\delta r)^2$, whereas 
for the $\beta$-dependent component we provide the scaling arguments (Methods) showing that at $\delta r=0$ the excitations are suppressed with decreasing $\beta$ too slowly to be completely eliminated in our numerical simulations. At  fixed $\beta$, this $\beta$-dependent contribution changes linearly with a small $\delta r$:
\begin{align}\label{nex_nad_NC}
  n_{\text{ex}}^{\text{nad}}(\delta r\slash g) -n_{\text{ex}}^{\text{nad}}(0)\propto  -\frac{\delta r}{g}.  
\end{align}
We plot the dependence of $n_{\text{ex}}^{\text{nad}}$ on $\delta r\slash g$ in the nonadiabatic regime in Fig.~\ref{fig3c}.  The numerical results in  Fig.~\ref{fig3b} were obtained by looking at the changes of $n_{ex}(\delta r)$ at the smallest possible but finite $\beta$. Hence, only the slope of the line (and not the constant offset from the origin) could be compared to Eq.~(\ref{nex-sat}).
\begin{figure*}
    \centering
    \labelphantom{fig3a}
    \labelphantom{fig3b}
    \labelphantom{fig3c}
    \includegraphics[width=\linewidth]{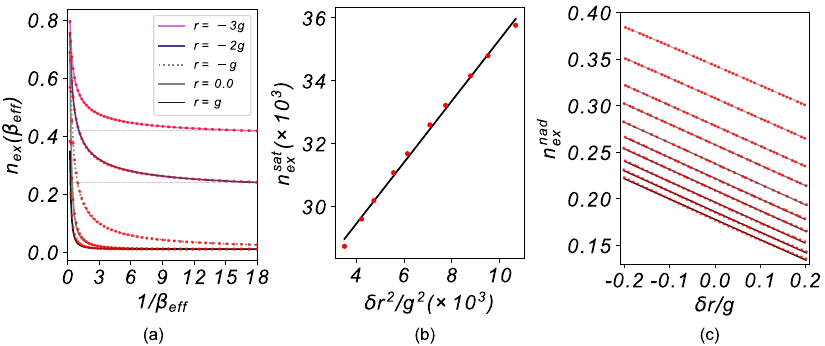}
    \caption{Dependence of the density of nonadiabatic excitations on the sweeping rate and interaction strengths (a) Numerically obtained dependence of the density $n_{\text{ex}}$ of the excitations on $\beta_{\rm eff}$ for different values of the coupling $r$ (at $N=100$). The curves correspond to the optimal fit for the exponent $\mu$ as predicted by \eqref{nex_tot}. (b) Asymptotic dependence of the quasi-adiabatic saturation value of $n_{\text{ex}}$ on the ratio $\delta r^2/g^2$ in the \textit{adiabatic} limit ($N=500$). The slope of the linear fit is ${\sim}0.97$, which corresponds well with the analytical expression in \eqref{nex-sat}.
    (c) Asymptotic dependence of the $\beta$-dependent part of $n_{\text{ex}}$  on  $\delta r/g$ for the non-adiabatic case at different values of sweeping rate $\beta_{\text{eff}}$, where $\beta_{\text{eff}}=(\beta\ln N)/g^2$ (decreasing from $1.5$ to $0.6$, top to bottom), the slopes of which are all the same (${\sim}0.22$) as expected ($N=200$).  }
    \label{fig3}
\end{figure*}

%\subsection{Touching the critical point}
%We also make a note on the regime that separates the two types of the phase transition that we already described. 

In general, for a finite sweeping rate $\beta$ at $\delta r=0$, we also have a Kibble-Zurek scaling of the number of excitations:
\begin{align} \label{KZmu}
    n_{{\text{ex}}}(\beta) \sim \beta^{\mu}
\end{align}
where the exponent $\mu$  depends on $\kappa$. In section~\ref{scaling}, we estimated this exponent analytically for $r=-g$ ($\kappa=-1$) to be $2\slash 3$, which is different from the exponent $1$ of the second-order phase transition (see Fig.~\ref{fig3a} for numerical proof). For $\kappa < -1$, we have numerically, together with \eqref{nex-sat}:
\begin{align}\label{nex_tot}
    n_{{\text{ex}}}(\beta) \sim C(\delta r)\beta^{\mu} + n_{\text{ex}}^{\text{sat}},
\end{align}
where $C(\delta r)$ is independent of $\beta$.

Previously, a similar scaling was observed for the formation of topological defects in the dynamics of a classical field undergoing a weakly first-order phase transition at finite temperature \cite{suzuki2023topological}. Here, we have obtained a similar result for a fully coherent evolution through a quantum phase transition at zero temperature.

\subsection{Transient dynamics and prethermalization}

%Need a gentler transition to this section. 
%After the discussion of transient oscillation, a clear elucidation of the relevance of prethermalization is in order - comparison at the level of quenches and creation of a 'lasting' nonequilibrium state whose lifetime can be parametrically tuned. 

%%%%%%%%%%%%%%%%%%%%%%%%%%%%

While the type of the phase transition can be understood using the scaling of the number of excitations obtained from the asymptotic number distribution of molecules, the transient dynamics, in the lead up to the number distribution as $t\rightarrow+\infty$, also exhibits features specific to the underlying phase transition. One characteristic feature accompanying the passage through the quantum critical point at time $t=0$ is the oscillation of atomic and molecular populations which become rapidly damped with time. Such transient oscillations of the number of molecules are usually observed in the nonadiabatic regime with a finite sweep rate $\beta>0$. In fact, the frequency dependence of such nonadiabatic oscillations between atomic and molecular populations was used in \cite{zhang2023many} to infer the role of \textit{bosonic} enhancement of the ultracold reaction. These oscillations are expected to vanish in the adiabatic limit $\beta\rightarrow 0$ as per the mean-field theory detailed in previous sections, as the dynamics is supposed to `equilibrate' to one of the minima of the mean-field potentials in \eqref{H_MF_IIPT} and \eqref{H_MF_IPT} in the $t\rightarrow\infty$ limit. 

 In order to understand this transient quantum dynamics and compare it with mean-field theoretic predictions, we plot the numerically obtained instantaneous average number of molecules computed as:
\begin{align}
    n_{\text{avg}}(t) = \frac{1}{N}\sum_{n=0}^N n \; | \langle n |\psi_{\text{mol}}(t)  \rangle|^2
\end{align}
for the quasi-adiabatic passage through a first-order (Fig.~\ref{fig4a}) and second-order (Fig.~\ref{fig4d}) quantum critical point. Here $\ket{\psi_{\text{mol}}}(t)$ is computed as $\ket{\psi_{\text{mol}}(t)} = e^{i\hat{H}t}\ket{\psi}_{n=0}$ where $\ket{\psi}_{n=0}$ corresponds to the initial state with no molecules and $\hat{H}$ is the Hamiltonian in \eqref{Htotal}. Numerically, this is implemented using a Trotter-factorized discretization of the unitary propagator $ e^{i\hat{H}t}$. (Since \eqref{Htotal} is also sparse, as is evident from \eqref{H_TC_matrix}, the time complexity of the cost of the propagation scales almost linearly with $N$.) 

For the case of a second-order phase transition, Fig.~\ref{fig4d} (plotted for $\kappa=0$) shows that $n_{\text{avg}}(t)$ increases monotonically with time to its saturation value of unity indicating a complete conversion of atoms to molecules as predicted by mean-field theory in the adiabatic limit. Fig. \ref{fig4f} also shows that the molecule number distribution $P(n_\text{mol})$ computed as:
\begin{align}
    P({n_\text{mol}}) = |\langle n_{\text{mol}} | \psi \rangle|^2
\end{align}
(where $\ket{\psi}$ is  the molecular wavefunction at various instances of time) becomes increasingly sharply peaked as $t\rightarrow\infty$.

However, we found that the transient state in the case of a first-order phase transition remains persistent and highly nonclassical. This is manifested as a complex series of \textit{collapse} and \textit{revival} of the number of molecules as a function of time (see Fig.~\ref{fig4a}). This beating pattern is particularly prominent in the adiabatic limit and is distinct from the non-adiabatic oscillations reported in \cite{zhang2023many}. We note that such patterns have been encountered previously in the  time-independent versions of the Jaynes-Cummings model \cite{shore1993jaynes}, Bose-Hubbard model \cite{milburn1997quantum}, a model of nonlinear directional couplers \cite{chefles1996quantum}, and even in a generalized version of the model in \eqref{H_TC} but for other starting conditions
\cite{santos2006classical}.

In stark contrast to the dynamics around a second-order critical point, the number distribution for the first-order transition is very broad even as $t\rightarrow\infty$, which indicates a phase squeezing. As discussed for related models in \cite{milburn1997quantum,greiner2002collapse}, the existence of such collapse-revival patterns is a purely quantum-mechanical phenomenon; the semiclassical Hamiltonian in \eqref{H_MF} does not predict these features.

\vspace{5pt}
\begin{figure*}
    \centering
    \labelphantom{fig4a}
    \labelphantom{fig4b}
    \labelphantom{fig4c}
    \labelphantom{fig4d}
    \labelphantom{fig4e}
    \labelphantom{fig4f}
    \labelphantom{fig4g}
    \labelphantom{fig4h}
    \includegraphics[width=\linewidth]{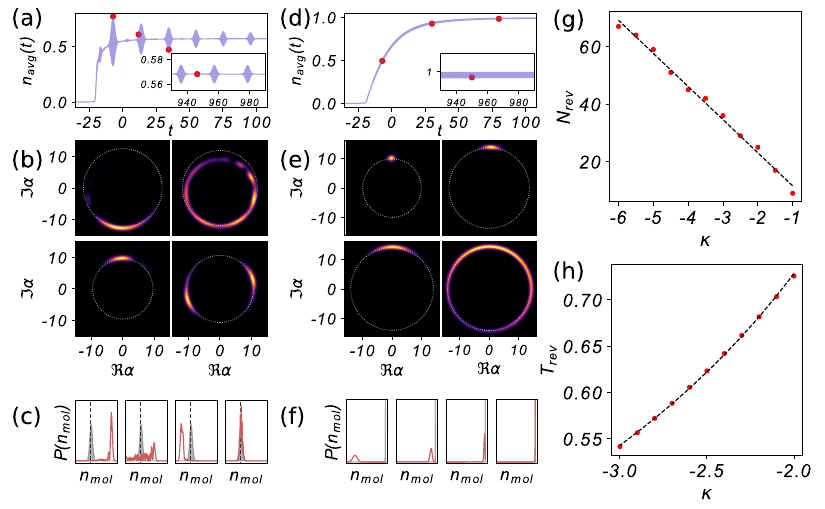}
    \caption{Characteristics of the quantum dynamics during the passage through a critical point (All the plots in this figure have been generated for $N=200$). (a) Plot of $\langle n_{\text{mol}}\rangle$ (scaled by $N$) vs time for $\kappa=-3$, i.e. for quasi-adiabatic first-order transition. Inset: Dependence of $\langle n_{\text{mol}}\rangle$ with $t$ in the long-time limit. (b) Coherent projection of the molecular wavefunction (see Methods) at the times marked in red dots in (a) (left-right-inset in (a), corresponds to left-right-top-bottom in (b)). (c) Plots of the number distribution of the molecules for the times marked in (a) (from left to right, followed by inset). The grey plot corresponds to the average asymptotic number distribution in the limit $t\rightarrow\infty$. (d) Same as in (a), but for the second-order phase transition at $\kappa=0$. (e) Same as in (b), but with $\kappa=0$. (f) Same as in (c), but with $\kappa=0$. (g) Plot of the number of the revivals in a given time window for different values of $\kappa$ in the first-order phase transition  (same $g$, adiabatic limit), fit to a linear curve (dashed line). (h) Time period of a particular revival as a function of $\kappa$ fit to $T_{\text{rev}} \sim 1\slash |\kappa|$ (dashed line). }
    \label{fig4}
\end{figure*}

To further understand the aspects of quantum dynamics that lead to this collapse-revival oscillation, we generated \textit{Husimi Q projections} of the time-dependent molecular wavefunction onto the basis of coherent states, as shown in Fig.~\ref{fig4e} (for second-order transition) and  Fig.~\ref{fig4b} (for first-order transition). The coherent Husimi Q projections of the wavefunction in the molecule number basis is obtained as:
\begin{align}
    Q(\alpha) &= \frac{1}{\pi} |\langle \alpha | \psi_{\text{mol}}\rangle|^2, \\
   \langle \alpha | \psi_{\text{mol}} \rangle &= e^{-|\alpha|^2\slash 2} \sum_{n=0}^N \frac{\alpha^{n}\psi_{\text{mol}}(n)}{\sqrt{n!}},
\end{align}
where $Q(\alpha)$ corresponds to the projection onto a coherent state indexed by $\ket{\alpha}$ and $\psi_{\text{mol}}(n)$ is the amplitude of the wavefunction in the molecular population basis $\ket{n}$ at any given time $t$.

The projections reveal that the collapse and revival patterns constitute an oscillation of the number distribution $P(n_{\text{mol}})$ about the asymptotic limit (marked in gray in Fig.~\ref{fig4c} and~\ref{fig4f}). For the second-order transition, since the distribution becomes increasingly peaked around $n_{\text{mol}}=1$ as $t\rightarrow\infty$, the phase of the molecular wavefunction $\ket{\psi_{\text{mol}}(t)}$ becomes increasingly diffused along a circle of radius $R$ ($\propto \sqrt{n}$, where $n$ is the instantaneous number of molecules) which is numerically demonstrated in Fig.~\ref{fig4e}. This is due to the number-phase uncertainty relation, as the molecular number distribution $P(n_{\text{mol}})$ becomes increasingly narrow as $t\rightarrow\infty$.

In the case of a first-order transition, the molecular number distribution at $t\rightarrow\infty$ is not sharp even in the adiabatic limit, which allows for the collapse-revival oscillation. In the Husimi projection, the `crest' and `trough' of the envelope of these oscillations localise on points along the diameter of a circle which differ in phase by $\pi$. They also correspond to an instantaneous peak of $P(n_{\text{mol}})$ above and below the asymptotic number distribution as shown in Fig.~\ref{fig4c}. The molecular wavefunction keeps `vacillating' between these two extremes until the $t\rightarrow\infty$ limit, when the trough and crest make way for the `cat state' in the molecular wavefunction,  as shown in Fig.~\ref{fig4b}.

We also found that the number $N_{\text{rev}}$ of revivals per unit time increases linearly with the strength of nonlinearity ($|\kappa|$) for the first-order transition. The frequency of the collapse-revival patterns decreases gradually following the sweep, and in general saturates to a constant value resulting in a punctuated series of revivals that survive for a very long time. The time-period $T_{\text{rev}}$ of any particular collapse and revival pair also decreases with $|\kappa|$ as
\begin{align}\label{Trev_vs_k}
    T_{\text{rev}} \sim \frac{1}{|\kappa|}.
\end{align}
We plot the numerically found dependence of $N_{\text{rev}}$ and $T_{\text{rev}}$ on $\kappa$ in Fig.~\ref{fig4g} and Fig.~\ref{fig4h} respectively. The sustained oscillation pattern and the dependence of the frequency and time period on the interaction strength are reminiscent of the \textit{prethermalized} states \cite{ueda2020quantum} that were numerically observed \cite{kollath2007quench,eckstein2009thermalization} following a sudden quench in bosonic/fermionic Hubbard models. A main aspect of such nonequilibrium states is the existence of long-time memory of the initial conditions as in our model. 

\vspace{5pt}
The formation of a quantum cat state is one of the main consequences of this long-term memory and inherently nonequilibrium dynamics of the prethermalized state. In the limit $t\rightarrow +\infty$, after the adiabatic first-order transition, the dynamics eventually freezes in a superposition of two macroscopically distinguishable states (shown as the bottom-right panel in Fig.~\ref{fig4b}) which differ by a phase of $\pi$. Each of these states in the Glauber basis is localized, and therefore is similar to a macroscopic Bose-Einstein condensate with a well defined phase. Thus, even the prethermalized state of the nonadiabatic excitations is strongly nonclassical.

\section{Discussion}
We have thus explored the role of interactions in mediating the passage through quantum phase transitions in ultracold chemistry with an extension of the Tavis-Cummings model. The standard integrable driven Tavis-Cummings model has made several predictions which should be observable in experiments that associate ultracold atoms into molecules by a stimulated passage through the Feshbach resonance. We tested the robustness of these predictions against other non-resonant interactions that break the model's integrability. For moderate interactions we found that, as in the integrable case, the system passes through the second-order quantum phase transition. In the quasi-adiabatic regime,  this leads to a nearly linear scaling, $n_{ex}\propto \beta \ln \beta$, of the number of the unformed molecules on the sweep rate $\beta$.
  
Above a certain critical interaction strength, the order of the phase transition changes from the second to the first, with drastic changes in the reaction dynamics. The main characteristic of the crossover to the first-order phase transition is that even in the adiabatic limit, the density of non-adiabatic excitations remains finite. Accompanying this is a transient dynamics that manifests as pronounced oscillations in the molecular population.  Asymptotically, the system reaches a prethermalized state (Fig.~\ref{fig4b}), which is a strongly nonclassical superposition of two condensate-like states with a phase difference of $\pi$ between them, a feature which is absent in the typical case of the second-order phase transition. Such dynamical features, which are usually attributed to the strongly nonadiabatic regime (e.g., following a sudden quench, as in \cite{kollath2007quench,eckstein2009thermalization,birnkammer2022prethermalization,ueda2020quantum,kaminishi2015entanglement}),  are observed in our model \eqref{Htotal}, during the quasi-adiabatic transition through a first-order critical point. 

Our analysis provides experimentally testable scaling laws for the number of nonadiabatic excitations as in \eqref{nex-sat},\eqref{nex_nad_NC} and \eqref{nex_tot} and for the time period of collapse-revival oscillations as in \eqref{Trev_vs_k}. The number of nonadiabatic excitations can indeed be experimentally measured as a function of the magnetic field ramp as was shown in (Fig.~4a of) \cite{zhang2021transition}. The coherent oscillations of atoms and molecules observed in \cite{zhang2023many} is a feature of the dynamics \textit{away} from the adiabatic limit, whereas the collapse-revival oscillation we present are most prominent in the adiabatic limit. The collapse-revival oscillations are indeed experimentally observable as was demonstrated in an experimental implementation of the Bose-Hubbard model using ultracold atoms in \cite{greiner2002collapse}. While the experimental observation of the cat state may indeed be challenging due to finite lifetimes of the atoms and molecules in a trap, as Fig.~\ref{fig4b} and \ref{fig4c} demonstrate, the features of the transient quantum dynamics are indicative of the ensuing cat state and may help design experiments to preserve the quantum coherence required for its formation.

The possibility of creating non-classical states such as the macroscopic cat states and squeezed states in bosonic systems has potential impact in quantum sensing and metrology \cite{pirandola2018advances}. In the past, several experiments have targeted the creation of such non-classical states \cite{kirchmair2013observation, ourjoumtsev2011observation, grimm2020stabilization}, some of which used a Kerr-type nonlinearity to stabilize these states. Given that the size of an atomic cloud is of the order of $N>10^4$, the passage through the Feshbach resonance with a first-order phase transition presents an attractive opportunity to create practically desired strongly nonclassical macroscopic states. 

In addition to applications in quantum sensing, experiments in ultracold chemistry also serve to demonstrate the universality of the dynamics around quantum phase transitions \cite{clark2016universal}. For instance, there is considerable evidence (both theoretical and observational) that phase transitions occurred during the early evolution of our universe \cite{cosmology-1}. Our study reveals that purely quantum correlations play a considerable role in such processes and that the nature of the quantum critical point is manifested in the scaling of the excitation density. The main experimental challenge in studying such correlations will be in tuning the interaction strength and obtaining the desired degree of quantum control. With major strides in experimental development in ultracold chemistry \cite{wang2024dynamics,zhang2021transition,zhang2023many} in the last decade, we expect this to be possible in the near future.

\section{Methods}

\subsection{Empirical models for ultracold reactions}\label{TC-bos}

We follow \cite{timmermans1999feshbach} in order to explain the relevance of simple empirical models (such as the Tavis-Cummings model) in describing ultracold reactions. The most fundamental model that describes all the many-body interactions in a gas of ultracold atoms is given as $\hat{H} = \int d\textbf{r} \:\hat{H}_{\text{a}}(\textbf{r})
$ where
\begin{equation}\label{Hatom}
\begin{split}
 \hat{H}_{\text{a}}(\textbf{r}) = \hat{\psi}_a^{\dag}(\textbf{r})\left( - \frac{\hbar^2\nabla^2}{2m_a} + V^a_{\text{ext}}(\textbf{r})\right) \hat{\psi}_a(\textbf{r})\\ + \frac{\lambda_{aa}}{2}\hat{\psi}_a^{\dag}(\textbf{r})\hat{\psi}_a^{\dag}(\textbf{r})\hat{\psi}_a(\textbf{r}) \hat{\psi}_a(\textbf{r}) 
\end{split}
\end{equation}
where $\lambda_{aa}$ is the (elastic) interaction strength between two atoms and $\hat{\psi}_a(\textbf{r})$ is the atomic field operator that destroys an atom at position $\textbf{r}$. This is written under the assumption of a short-range interaction described by the Fermi pseudopotential as:
\begin{align}\label{pseduopot}
    \lambda_{aa} = \frac{4\pi\hbar^2 a_{aa}}{m_a}
\end{align}
where $a_{aa}$ is the atom-atom scattering length. If the atoms are subject to a magnetic field, in the vicinity of a Feshbach resonance, the scattering length varies as $a_{aa}(B) = a_{aa}(1-\Delta B\slash (B-B_m))$ where $B_m$ is the peak of the resonance and $\Delta B$ is the width, which changes the interaction strength $\lambda_{aa}$ accordingly. However, this approach fails to describe the transformation of atoms into molecules at the resonance, so the scattering problem needs to be treated by considering molecules as separate compound particles described by the field operator $\hat{\psi}_m(\textbf{r})$. 

Then the resonant interaction between atoms and molecules is computed as a sum of all state-to-state scattering matrix elements of the hyperfine interaction $\bra{\textbf{K}|\hat{V}_{\text{hf}}} \ket{\textbf{k},\textbf{k'}}$, where $\ket{\textbf{k}},\ket{\textbf{k}'}$ corresponds to the atomic field and $\ket{\textbf{K}}$ to a molecular field with the respective momentum vectors (see eq.~(46) to (49) in \cite{timmermans1999feshbach}). Under the assumption of an \textit{elastic} interaction, the resonant term of the atom-molecule Hamiltonian becomes:
\begin{align}\label{Ham_res}
    \hat{H}_{\text{am}}^{\text{res}} =  \alpha \int d \textbf{r} \: (\hat{\psi}^{\dag}_m(\textbf{r})\hat{\psi}_a(\textbf{r})\hat{\psi}_a(\textbf{r}) + \hat{\psi}_m(\textbf{r})\hat{\psi}^{\dag}_a(\textbf{r})\hat{\psi}^{\dag}_a(\textbf{r}))
\end{align}
In line with \eqref{Hatom}, the Hamiltonian for the compound atom+molecule system ought to contain all elastic interactions thereby yielding $\hat{H}_{\text{all}} = \int d\textbf{r} \:\hat{H}_{\text{all}}(\textbf{r})$ where:
\begin{equation}
\begin{split}
\hat{H}_{\text{all}}(\textbf{r}) =\:&  \hat{\psi}_a^{\dag}(\textbf{r})\left( - \frac{\hbar^2\nabla^2}{2m_a} + V_{\text{ext}}^a(\textbf{r}) \right) \hat{\psi}_a(\textbf{r}) \:+ \\& \frac{\lambda_{aa}}{2}\hat{\psi}_a^{\dag}(\textbf{r})\hat{\psi}_a^{\dag}(\textbf{r})\hat{\psi}_a(\textbf{r}) \hat{\psi}_a(\textbf{r}) \: + \\ &\hat{\psi}_m^{\dag}(\textbf{r})\left( - \frac{\hbar^2\nabla^2}{2m_m} + V^m_{\text{ext}}(\textbf{r}) \right) \hat{\psi}_m(\textbf{r}) \:+ \\ &\frac{\lambda_{mm}}{2}\hat{\psi}_m^{\dag}(\textbf{r})\hat{\psi}_m^{\dag}(\textbf{r})\hat{\psi}_m(\textbf{r}) \hat{\psi}_m(\textbf{r}) \:+  \\
&\lambda_{am}\hat{\psi}_a^{\dag}(\textbf{r})\hat{\psi}_a(\textbf{r}) \hat{\psi}^{\dag}_m(\textbf{r})\hat{\psi}_m(\textbf{r}) \:+ \\ & \alpha (\hat{\psi}^{\dag}_m(\textbf{r})\hat{\psi}_a(\textbf{r})\hat{\psi}_a(\textbf{r}) + h.c.)
\end{split}
\end{equation}
where $\lambda_{am}$ and $\lambda_{mm}$ are atom-molecule and molecule-molecule scattering length. This is the \textit{minimal} Hamiltonian that ought to be considered to include all interactions that come from (resonant and non-resonant) elastic scattering between atoms and molecules. 

In the low-temperature limit, the kinetic energy terms are insignificant, and spatial homogeneity of the field operators is a fair assumption. Assuming a linear ramp of the magnetic field at the rate $\beta$ (which forms the $V_{\text{ext}}$ term), the resultant time-dependent Hamiltonian is given as:
\begin{equation}\label{H_all_main}
\begin{split}
    \hat{H}_{\text{all}}(t) =\: & \frac{\beta t}{2}\hat{\psi}_a^{\dag}\hat{\psi}_a - \beta t \:\hat{\psi}_m^{\dag}\hat{\psi}_m + \alpha (\hat{\psi}_m^{\dag}\hat{\psi}_a\hat{\psi}_a + \hat{\psi}_m\hat{\psi}_a^{\dag}\hat{\psi}_a^{\dag}) \:  \\
    & +\frac{\lambda_{aa}}{2}\hat{\psi}_a^{\dag}\hat{\psi}_a^{\dag}\hat{\psi}_a \hat{\psi}_a + \frac{\lambda_{mm}}{2}\hat{\psi}_m^{\dag}\hat{\psi}_m^{\dag}\hat{\psi}_m \hat{\psi}_m \\ &+ \lambda_{am}\hat{\psi}_a^{\dag}\hat{\psi}_a \hat{\psi}_m^{\dag}\hat{\psi}_m
\end{split} 
\end{equation}
Truncating the above equation upto the terms in the first line yields the model Hamiltonian considered in \cite{zhang2021transition} (the sign of $\beta$ depends on the starting state, but the time-dependant terms for atom and molecule come with opposite signs always). However, we emphasize that there is no theoretical justification for doing the same.

\subsection{Relating Tavis-Cummings models to other models of molecular dissociation}\label{TC-mol}

In the derivation of \eqref{H_all_main}, the component atoms are assumed to be of the same `type' but that can be easily adjusted. For example, consider the reaction of bosons $\Psi \leftrightarrow a+b$, in which $\hat{\Psi}$, $\hat{a}$, $\hat{b}$ are all bosonic annihilation operators. The Hamiltonian $\hat{H}_{\text{all}}(t)$ becomes (in the notation of \eqref{H_TC}):
\begin{eqnarray}
\label{H11}
 \hat{H}_B(t) :=
 %-\frac{\beta t}{2} (\hat{a}^{\dagger}\hat{a} + \hat{b}^{\dagger}\hat{b}) + 
  \beta t \hat{\Psi}^{\dagger} \hat{\Psi}+\frac{g}{\sqrt{N}}\left(\hat{\Psi}^{\dagger}\hat{a}\hat{b}  +\hat{\Psi}\hat{a}^{\dagger}\hat{b}^{\dagger} \right).
\end{eqnarray}
where we absorb the time-dependent term that describes interactions between atoms and magnetic field into the respective term for molecular interaction. Since one molecule splits into two atoms, and the number of atoms of $a$ and $b$ type is the same, the time evolution of \eqref{H11} conserves the number 
\begin{align}
    \hat{N} \equiv \hat{\Psi}^{\dag}\hat{\Psi} + (\hat{a}^{\dagger}\hat{a}+\hat{b}^{\dagger}\hat{b})/2,
\end{align}
whose eigenvalue $N$ corresponds to the initial conditions with $N$ molecules. Hence, it is convenient to mark all states  by the number of split molecules, $n$: 
\begin{equation}
|n\rangle \equiv |N-n\rangle_m \otimes |n\rangle_{a} \otimes |n\rangle_b,
\label{basis-b}
\end{equation}
where, e. g., $\hat{a}^{\dagger}\hat{a}|n\rangle_a=n|n\rangle_a$.
The initial state with $N$ molecules as $t\rightarrow -\infty$ corresponds to $|n\rangle =|0\rangle$. It is then straightforward to see that the only non-zero off-diagonal matrix elements of the Hamiltonian are:
\begin{align}
    \langle m+1|\hat{H}_B(t)|m\rangle =(N-m)\sqrt{(m+1)}.
\end{align}
%and its corresponding superdiagonal element. 
Comparing with (\ref{H11}), we find  
$$
\langle m|H|m'\rangle=\langle m|H_{TC}|m'\rangle, \quad \forall \, m,m'.
$$

Another related example is that of a pair of  fermions (with opposite spin) transforming into a boson: $\psi \leftrightarrow a_{\uparrow}+a_{\downarrow}$. We can describe this reaction by using the following pseudo-spin operators (the pseudo-spin description is only useful for fermion-boson transformation) as in \cite{altland2009nonadiabaticity}:
\begin{align}\label{pseudospin}
    \hat{S}^+ = \hat{a}^{\dag}_{\downarrow}\hat{a}^{\dag}_{\uparrow} \;\;\;\: \hat{S}^- = \hat{a}_{\downarrow}\hat{a}_{\uparrow} \;\;\;\: \hat{S}_z = \frac{1}{2}(  \hat{a}^{\dag}_{\downarrow}\hat{a}_{\downarrow}+\hat{a}^{\dag}_{\uparrow}\hat{a}_{\uparrow})
\end{align}
where $\hat{a}^{\dag}_{\uparrow\slash\downarrow}$ and $\hat{a}_{\uparrow\slash\downarrow}$ are the fermionic creation and annihilation operators in the up/down spin state. 

Again, truncating upto the terms in the first line of \eqref{H_all_main} and using the number conservation property, we get the driven Tavis-Cummings model in \eqref{H_TC} which is the starting model of this article:
\begin{align}
    \hat{H}_{TC}(t) = -\beta t \hat{\psi}^{\dagger} \hat{\psi} + \frac{g}{\sqrt{N}}(\hat{\psi}^{\dagger} \hat{S}^{-}+\hat{\psi} \hat{S}^+)
\end{align}
Thus, both bosonic and fermionic ultracold reactions can be analysed using the same theoretical tools described in this article. Also, since the resonant transformation of atom to molecules is given by the term 
\begin{align}
    \hat{H}_{\text{am}}^{\text{res}} = g( \hat{\psi}^{\dag}\hat{S}^- + \hat{\psi}\hat{S}^+),
\end{align}
in the TC model, the non-resonant atom-molecule coupling term is given as:
\begin{align}
    \hat{H}_{\text{am}} =  r_{\text{am}}\:\hat{n}\hat{S}_z
\end{align}
the non-resonant atom-atom coupling term as:
\begin{align}
    \hat{H}_{\text{aa}} =  r_{\text{aa}}\hat{S}^+\hat{S}^-
\end{align}
(that is, outside the resonance, a pair of atoms which are `destroyed' form a new pair of atoms instead of molecules) and the molecule-molecule coupling term as:
\begin{align}
    \hat{H}_{\text{mm}} =  r_{\text{mm}}\hat{n}^2
\end{align}
By the number conservation property $\hat{n}+\hat{S}_z = N$, the terms proportional to $\hat{n}^2$ can be collected to yield:
\begin{align}
    r := r_{\text{mm}} - r_{\text{aa}} - r_{\text{am}}
\end{align}
This makes sense physically because, if a repulsive interaction between molecules increases the efficiency of the Feshbach transformation, then repulsive interaction between atoms (or between atoms and molecules) should have the opposite effect. The terms proportional to $\hat{n}$ only lead to a time shift in the resultant dynamics.

\subsection{Higher order Feshbach resonances}
{In Ref.~\cite{zhang2023many}, a different type of Feshbach resonance was experimentally studied: for reaction $c+\Psi \leftrightarrow a+b+c$, where $a,b,c$ are bosonic atoms. In our notation, this corresponds to making the reaction amplitude, $g$, dependent on the number of atoms, i.e., assuming in Eq.~(\ref{H11}) that $g\rightarrow gn/\sqrt{N}$, where $n$ is now the number of the \textit{decayed} molecules. We again consider the additional interaction term in the form $H_{int}=r\hat{n}^2/N$, where in the basis (\ref{basis-b}), we can define $\hat{n}=(\hat{a}^{\dagger}\hat{a}+\hat{b}^{\dagger}\hat{b})/2$. The entire bosonic Hamiltonian is then 
$$
\hat{H}=\beta t \hat{\Psi}^{\dagger} \hat{\Psi} +\frac{r\hat{n}^2}{N}+ \frac{g\hat{n}}{N}\left(\hat{\Psi}^{\dagger}\hat{a}\hat{b}  +\hat{\Psi}\hat{a}^{\dagger}\hat{b}^{\dagger} \right)
$$
(the correspondence between the model above and the model considered in \cite{zhang2023many} can easily be seen by setting $\hat{a}=\hat{b}$, i.e. considering identical atoms). The matrix elements $H_{nm}\equiv \langle n| \hat{H} |m\rangle$ in the basis (\ref{basis-b}) are given by 
\begin{equation}\label{H_TC_matrix-4}
\begin{split}
H_{nm}=& \:(-\beta t n+\frac{r}{N}n^2)\: \delta_{n,m} +  g n^2 \sqrt{\frac{N-n+1}{N}} \:\delta_{n,m-1}   \\&+ \: g(n+1)^2 \sqrt{\frac{N-n}{N}} \:\delta_{n,m+1}
\end{split}
\end{equation}

Repeating our derivation of the semiclassical Hamiltonian (\ref{eff-H2}) at low $n$,
we find a classical Hamiltonian, in which the saddle point again corresponds to $\phi=\pi$. Performing then the same canonical transformation (\ref{can_trans}) for the case of  $1\ll n\ll N$, we identify the
 potential energy by setting the momentum to zero, $P=0$, in the classical Hamiltonian, which gives us 
\begin{align}
   \label{eff-H4}
    V_{\rm eff}(Q) = -\beta t \: Q^2 + \left(\frac{r+g}{N}\right)Q^4 + \ldots,
\end{align}
The main difference from the case of the Tavis-Cummings model is the lack of the renormalization of the 
chemical potentials by the interaction term. Now, there is no $\propto g Q^2$ contribution to the Hamiltonian from the interactions, so the critical point is passed precisely at $t=0$. As in the Tavis-Cummings model, either the 2nd order, for $r+g>0$, or the 1st order, for $r+g<0$ phase transition is  then expected at the Feshbach resonance. Thus the presence of non-resonant interactions considered in this work is resilient to the presence of three-body recombination events studied in \cite{zhang2023many}. The critical exponents, however, are expected to be different because of a more complicated kinetic energy term in the classical Hamiltonian that corresponds to Eq.~(\ref{H_TC_matrix-4}).}

\subsection{Scaling exponent beyond Painleve-II equation}\label{scaling}
When a Hamiltonian has a time-dependent perturbation $\gamma$, the nonadiabatic tunneling probability is computed from the change in the curvature (or frequency) $\omega_*$ \cite{landau1982mechanics} as:
\begin{align}\label{Gamma_NA}
    \Gamma = \frac{\Delta I}{2}  :=\Re \int_{-\infty}^{\infty}  i e^{i\theta}  \frac{ \dot{\gamma} d\theta}{\omega_*(I,\gamma)}
\end{align}
Following the approach in   \cite{liu2002theory}, to evaluate the scaling behaviour of $\Gamma$, we  express $\omega_*$ in terms of $\theta$. In the case of the first-order transition, the major change in the curvature occurs at the point $t=t_*$ (see \eqref{timeshift}) after which $n$ increases in time. We consider the dynamical equations given by the Hamiltonian \eqref{H_MF} in terms of the scaled time variable
$\tau = g t$ which yields:
\begin{align}
  \frac{d\overline{n}}{d\tau} &= 2 \overline{n}\sqrt{1 -\overline{n}} \sin \phi := f_1(\overline{n},\phi) \\
    \frac{d\phi}{d\tau} &= \overline{\gamma} + 2\kappa \overline{n} + {\cos \phi} \left(\frac{2-3\overline{n}}{\sqrt{1-\overline{n}}} \right)  :=f_2(\overline{n},\phi)
\end{align}
where $\overline{\gamma} = {\gamma}\slash{g},
\kappa = {r}\slash{g}$ and $\overline{n} = {n}\slash{N}$. It is straightforward to see that the rate of change of $\overline{\gamma}$ is given as:
\begin{align}
    \frac{d\overline{\gamma}}{d\tau} = \frac{\beta}{g^2} := \frac{\beta_{\text{eff}}}{\log_e N}
\end{align} 
The reason for the inclusion of a logarithmic term in the above expression is explained in \cite{malla2022coherent}: for a given $\beta$, the $f =2\pi\slash \beta_{\text{eff}}$ is the point of discontinuity in the behaviour of $n_{\text{ex}}$ vs $\beta$ (see Eq.~(13) in \cite{malla2022coherent} and note that the parameter $g$ in \cite{malla2022coherent} differs from the $g$ in this article by a factor of $\sqrt{N}$).

\vspace{5pt}

Around the fixed point $(\overline{n}_*,\phi_*) =(0,\pi)$, for $\kappa=-1$, this yields:
\begin{align}
    \omega_* \sim \overline{n}_*
\end{align}
from which we get:
\begin{align}
     \omega_* \sim \beta_{\text{eff}}^{1\slash3} \theta^{1\slash3}
\end{align}
This gives the following expressions:
\begin{align}
   \Gamma_c  &\sim -\beta_{\text{eff}}^{\sfrac{2}{3}} \\
    \Rightarrow n_{\text{ex}}(\beta_{\text{eff}}) &\sim \beta_{\text{eff}}^{\sfrac{2}{3}} 
\end{align}
which we  confirmed numerically in Fig.~\ref{fig3a}. In general, we found that this power-law scaling holds for all values of $\kappa$, but an analytical derivation of the corresponding exponent $\mu$ was not possible.

\vspace{5pt}
For the near-critical case with $\kappa = -1+\delta \kappa$, we get:
\begin{align}
    \omega_* \sim  \left( \overline{n}_*+\frac{1}{3} {\delta \kappa} \right) 
\end{align}
Expanding upto linear order in $\kappa$, we get:
\begin{align}
    \Gamma = \Gamma_c + \Delta \Gamma_c
\end{align}
where 
\begin{align}
    \Delta \Gamma_c \sim \delta \kappa 
\end{align}
(we numerically found a weak dependence of $\Delta \Gamma_c$ with respect to $\beta_{\text{eff}}$). Note that the sign of $\Gamma_c$ is positive: when $\kappa$ increases, the tunneling probability increases (which can be inferred from fig. \ref{fig1c}. For a constant $g$, we hence have:
\begin{align}
    n_{\text{ex}}(\delta r)- n_{\text{ex}}(0)\propto -\frac{\delta r}{g}.
\end{align}
Thus, the concentration of defective excitations increases linearly with $1\slash g$, which is also evident from Figs.~\ref{fig1c} and \ref{fig3c}. 

\section{Data availability}
The simulation data generated in this study have been deposited \href{https://doi.org/10.17863/CAM.113114}{here} in the University of Cambridge's Apollo repository under the DOI: 10.17863/CAM.113114.

\section{Code availability}
The code used to generate the data in this study is also available in the \href{https://doi.org/10.17863/CAM.113114}{same} repository as the simulation data, under the same DOI: 10.17863/CAM.113114.

%apsrev4-2.bst 2019-01-14 (MD) hand-edited version of apsrev4-1.bst
%Control: key (0)
%Control: author (8) initials jnrlst
%Control: editor formatted (1) identically to author
%Control: production of article title (0) allowed
%Control: page (0) single
%Control: year (1) truncated
%Control: production of eprint (0) enabled
%
%\bibliography{sn-bibliography}% common bib file
%% if required, the content of .bbl file can be included here once bbl is generated
%%\input sn-article.bbl

\section{Acknowledgement }
This work was supported in part by the U.S. Department of Energy, Office of Science, Office of Advanced Scientific Computing Research, through the Quantum Internet to Accelerate Scientific Discovery Program, and in part by the U.S. Department of Energy under the LDRD program at Los Alamos. V.G.S. acknowledges funding from St. John's College, Cambridge for travel, and the Yusuf Hamied Department of Chemistry, Cambridge for computing resources. F.S. acknowledges support from the Center for Nonlinear Studies. We also thank Yair Litman and Paramvir Ahlawat for their scientific comments that improved the presentation style of the manuscript.

\section{Author contribution}
N.S. proposed the project. N.S. and V.G.S. designed the research; V.G.S. and F.S. performed the research; B.Y. analysed data and edited the paper; V.G.S. and N.S. wrote the paper. 

\section{Competing interests}
The authors declare no competing interests.

\end{document}